\documentclass[ prl, twocolumn, amsfonts, amsmath,floatfix] {revtex4-1}
\pdfoutput=1
\usepackage{graphicx}
\usepackage{amssymb,amsmath}
\usepackage[usenames, dvipsnames]{color}
\usepackage{soul}
\usepackage{lipsum}
\usepackage{natbib}
\usepackage[mathscr]{eucal}
\usepackage{dsfont}

\renewcommand{\vec}[1]{{\mathbf #1}}
\newcommand{\av}[1]{\langle #1\rangle}
\newcommand{\mc}[1]{{\mathcal #1}}
\newcommand{\be}{\begin{equation}}
\newcommand{\ee}{\end{equation}}
\newcommand{\str}[1]{{\textrm #1}}
\usepackage[usenames,dvipsnames]{color}

\begin{document}

\title{Mutual information between reflected and transmitted speckle images}

\author{
 N. Fayard, A. Goetschy, R. Pierrat, and R. Carminati,
}

\affiliation{
ESPCI Paris, PSL Research University, CNRS, Institut Langevin, 1 rue Jussieu, F-75005 Paris, France
}

\begin{abstract}

We study theoretically the mutual information between reflected and transmitted speckle patterns produced by wave scattering from disordered media. The mutual information between the two speckle images recorded on an array of $N$ detection points (pixels) takes the form of long-range intensity correlation loops, that we evaluate explicitly as a function of the disorder strength and the Thouless number $g$. Our analysis, supported by extensive numerical simulations, reveals a competing effect of cross-sample and surface spatial correlations. An optimal distance between pixels is proven to exist, that enhances the mutual information by a factor $Ng$ compared to the single-pixel scenario.

\end{abstract}

\maketitle

When waves propagate in complex environments, their information content is spread out in space and encoded into complicated speckle patterns, eventually recorded as two-dimensional images at the output of the medium. A central issue is the quantification of the information content in speckle patterns, and its use for image recovery, power deposit, or information delivery \cite{mosk12, rotter17, moustakas00}. In this perspective, much effort has been made in recent years to take advantage of the existence of spatial correlations in speckles measured in transmission. Various schemes based on the memory effect of short-range correlations (termed $C^{TT}_1$ hereafter), have been developed to image an object placed behind an opaque screen \cite{bertolotti12, katz14}, while long-range correlations ($C^{TT}_2$), that capture nonlocal information, have been demonstrated to be useful for increasing energy delivery through turbid media \cite{popoff14a, hsu17}. 

Very recently, the existence of cross-correlations between speckle patterns measured in reflection and transmission has been demonstrated, and the shape of the intensity correlation function has been characterized in regimes ranging from quasi-ballistic to diffusive transport~\cite{fayard15, starshynov17}. The presence of  correlations suggests the possibility to acquire some information about a transmitted speckle based on a measurement restricted to the reflection half-space. This is of crucial importance for sensing, imaging, and communicating through turbid media, and for the control of wave transmission through disordered scattering environments by wavefront shaping techniques \cite{mosk12, rotter17, simon01}.  In this Letter, we quantify the amount of mutual information (MI) between transmitted and reflected speckles, and analyze the dependance of the MI on the disorder strength and the geometrical parameters characterizing the detection process (number of detectors and their interdistance). 

\begin{figure}[t]
\includegraphics[width=1\linewidth]{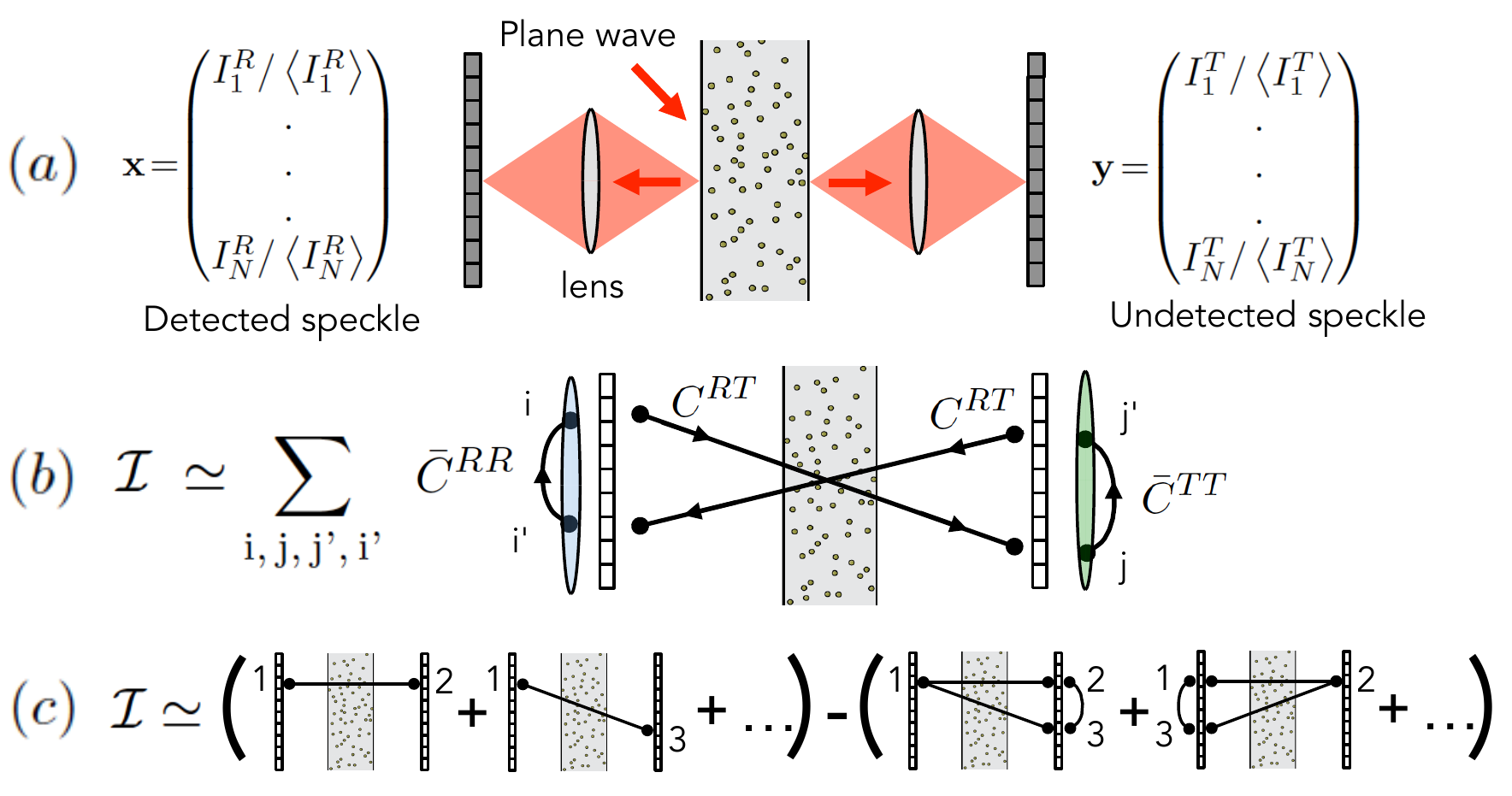}
\caption{(a) Schematic view of a disordered slab illuminated by a plane wave. The reflected speckle produced at the sample surface, $\vec{x}$, is registered on a CCD camera with $N$ pixels, and its mutual information $\mc{I}$ with the transmitted speckle $\vec{y}$ is evaluated. (b) Diagrammatic representation of the trace formula (\ref{InfoTrace}) as a sum of correlation loops. (c) Diagrammatic representation of the approximation (\ref{InfoTrace2}).} 
\label{fig1}
\end{figure}

The scheme of the \textit{gedanken} experiment is represented in Fig.~\ref{fig1}(a). A slab of a disordered medium is illuminated by a plane wave, and the speckle intensity profile is recorded with a CCD camera placed at the frontside. The transmitted speckle, potentially recorded with another camera, is assumed to be unknown. Let $I^R_i=\vert E_i\vert^2$ be the reflected intensity measured on pixel $i$ (or detector $i$) and $x_i=I^R_i/\av{I^R_i}$ the normalized intensity, the brackets $\av{\dots}$ denoting an ensemble average over statistical realizations of the disordered medium. The reflected speckle image is represented by the vector $\vec{x}$ of size $N$ equal to the number of pixels of the camera. Similarly, the transmitted unknown image is wrapped up into a vector $\vec{y}$. In a statistical description of the disordered medium, configurations of disorder are generated by a stochastic process, and $\vec{x}$ and $\vec{y}$ are random variables. A quantitative estimate of the statistical dependance between $\vec{x}$ and $\vec{y}$, or equivalently between the two speckle images, is given by their MI, defined as the difference between the entropy of $\vec{x}$ and $\vec{y}$ considered separately and the entropy of the pair $\{\vec{x},\vec{y}\}$ \cite{cover91}:
\be
\mc{I}=\iint \str{d}\vec{x}\str{d}\vec{y}\,p(\vec{x}, \vec{y})
\log
\left[\frac{p(\vec{x}, \vec{y})}{p(\vec{x})p(\vec{y})}\right].
\label{def_info}
\ee
Here $p(\vec{x})$, $p(\vec{y})$, and $p(\vec{x}, \vec{y})$ are joint probability density functions (PDF). Note the important difference with Multiple-Input Multiple-Output protocols, in which the MI between input and output signals is evaluated \cite{moustakas00, tulino04, skipetrov03, goetschy13_prl}. In our thought experiment, the input signal is not random, there is no external noise, and $\vec{x}$ is not the injected signal but the output signal in reflection.

One difficulty in evaluating the MI (\ref{def_info}) lies in the fact that the PDF $p(\vec{x})$, $p(\vec{y})$ and $p(\vec{x}, \vec{y})$ are theoretically unknown. Only marginal distributions, such as $p(x_1)$ and  $p(y_1)$, as well as two-point correlations functions ($C^{RR}_{ii'}=\av{\delta x_i \delta x_{i'}}$, $C^{TT}_{jj'}=\av{\delta y_j \delta y_{j'}}$, and $C^{RT}_{ij}=\av{\delta x_i \delta y_j}$, with $\delta x =x-1$ and $\delta y =y-1$) have been calculated for disordered media \cite{akkermans07, vanrossum99, starshynov17}. 
In the limit of small pairwise correlations however, we will show that $\mc{I}$ can be expressed as a combination of the previous correlators only, even if the field amplitudes $E_i$ cannot be modelled as complex Gaussian random variables. 

First, we express $p(\vec{x}, \vec{y})$ in terms of $p(\vec{x})$ and $p(\vec{y})$. This joint PDF is entirely characterized by the set of correlators
$\av{x^{\{n\}}y^{\{m\}}}=
\av{
x_1^{n_1}...x_N^{n_N}y_1^{m_1}...y_N^{m_N}
\label{TotMomentum}
}.
$
Since $\vec{x}$ and $\vec{y}$ are weakly correlated in the multiple scattering regime \cite{fayard15, starshynov17}, we search for leading corrections to the independent variable result $\av{x^{\{n\}}y^{\{m\}}}=\av{x^{\{n\}}} \av{y^{\{m\}}}$. To proceed, we adopt a path-integral-type representation of the field, expressing $E_i$ as a sum of propagators along all possible scattering trajectories $\mc{S}$ inside the medium, $ E_i=\sum_{\mc{S}}E_i^{\mc{S}}$. Hence, each term $x_i^{n_i}\varpropto\vert E_i \vert^{2n_i}$ contains $n_i$ replica of complex propagators $E_i^{\mc{S}}$ and $E_i^{\mc{S}*}$. The leading correction to the independent result involves all combinations of correlations between two reflection propagators and two transmission propagators. Counting these combinations yields
$\av{x^{\{n\}}y^{\{m\}}}= \av{x^{\{n\}}} \av{y^{\{m\}}}
+ \sum_{i,j}n_i^2n_j^2 \, \av{\delta x_i \delta y_j}\av{x^{\{n\}-1_i}}  \av{y^{\{m\}-1_j}}$,
where the notation $x^{\{n\}-1_i}=x_1^{n_1}\dots x^{n_i-1}\dots x_N^{n_N}$ is used. This expression gives the moments of $p(\vec{x},\vec{y})$ in terms of the moments of $p(\vec{x})$ and $p(\vec{y})$. Then, standard algebra, detailed in the Supplementary Information (SI) \cite{SI}, allows us to cast the joint PDF in the form $p(\vec{x},\vec{y})=p(\vec{x})p(\vec{y})[1+\sum_{i,j}u_{ij}(\vec{x},\vec{y})]$, with 
\be
u_{ij}(\vec{x},\vec{y})=\av{\delta x_i \delta y_j} 
\frac{\partial_{x_i}[x_i \partial_{x_i}p(\vec{x})]}
{p(\vec{x})}
\frac{\partial_{y_j}[y_j \partial_{y_j}p(\vec{y})]}
{p(\vec{y})}.
\ee

Second, we insert the previous decomposition into Eq.~(\ref{def_info}), and 
express the logarithm as a power series of the correlation function $\sum_{ij}u_{ij}(\vec{x},\vec{y})$. By keeping the first non-zero term in the power expansion, we obtain the following trace formula  \cite{SI}:
\be
\mc{I}\simeq \frac{1}{2\textrm{ln}2}\textrm{Tr}\left[
C^{RT}\bar{C}^{TT}C^{RT}\bar{C}^{RR}
\right].
\label{InfoTrace}
\ee 
In this expression we have introduced three $N\times N$ matrices, with elements defined as $C^{RT}_{ij}=\av{\delta x_i \delta y_j}$, 
\be
\bar{C}^{TT}_{jj'}=\int \str{d}\vec{y} \, \frac{ \partial_{y_j}[y_j \partial_{y_j}p(\vec{y})]\partial_{y_{j'}}[y_{j'} \partial_{y_{j'}}p(\vec{y})]} {p(\vec{y})},
\label{DefCbar}
\ee
and $\bar{C}^{RR}_{ii'}$ in which $p(\vec{y})$ is replaced by  $p(\vec{x})$. Equation~(\ref{InfoTrace}) has a clear interpretation: the MI between the reflected and transmitted speckle patterns is the sum of all correlation loops $(i \to j \to j' \to i' \to i)$, as illustrated in Fig.~\ref{fig1}(b). In each loop, the correlation between pixels in different images is carried by pairwise cross-sample long-range coupling ($C^{RT}_{ij}$ and $C^{RT}_{j'i'}$), whereas the correlations within each image ($\bar{C}^{RR}_{i'i}$ and $\bar{C}^{TT}_{jj'}$) are more complicated since they are nonlocal, involving the full distributions $p(\vec{x})$ and $p(\vec{y})$.

To make the interpretation of the trace formula (\ref{InfoTrace}) even more transparent, we further assume that the distance between pixels in each image is larger than the free-space wavelength $\lambda$, so that correlations within each image remain small. As detailed in the SI, this allows us to approximate the transmission PDF as  $p(\vec{y})=\prod_k p(y_k) [1+\sum_{j<j'}u_{jj'}(y_j,y_{j'})]$, where $p(y_k)=e^{-y_k}[1+C_2^{TT}(y^2/4-y+1/2)]$. Here $C_2^{TT}\simeq\av{(\delta y)^2}-1$ is the leading non-Gaussian local correction to the usual Rayleigh statistics \cite{schnerb91, kogan93}. The reflection side PDF $p(\vec{x})$ takes the same functional form, with $C_2^{RR}$ replacing $C_2^{TT}$. With this simplification, the matrix elements (\ref{DefCbar}) reduce to 
\be
\bar{C}^{TT}_{jj'}=(1-C_2^{TT})\delta_{jj'}-(1-2C_2^{TT})\av{\delta y_j \delta y_{j'}}(1-\delta_{jj'}).
\label{Cbar}
\ee
This result is a first order expansion in $C_2^{TT}$, that can be generalized to higher order if needed, as discussed in the SI. However, if we operate in a regime where local correction $C_2^{TT}$ and $C_2^{RR}$ are much smaller than unity, we simply get $\bar{C}^{TT}=\mathds{1}-C^{TT}$ and  $\bar{C}^{RR}=\mathds{1}-C^{RR}$, where the diagonal elements of the matrices $C^{TT}$ and $C^{RR}$ are zero. In that case, the trace formula (\ref{InfoTrace}) is well approximated by
\be
\mc{I} \simeq 
\frac{1}{2\textrm{ln}2}\textrm{Tr}\left[(\mathds{1}-{C}^{TT}-{C}^{RR})
(C^{RT})^2
\right].
\label{InfoTrace2}
\ee 
Hence, the existence of pairwise long-range correlations inside each image tends to reduce the MI between the two images compared to the result without surface correlation, $\textrm{Tr}\left[(C^{RT})^2\right]/2\textrm{ln}2$ [see Fig.~\ref{fig1}(c) for an illustration]. This means that long-range cross-sample correlations and long-range surface correlations compete each other, suggesting that a balance can be found that maximizes the MI for certain geometrical configurations of detectors. This effect is analyzed at the end of this Letter.

\begin{figure*}[t]
\includegraphics[width=0.8\linewidth]{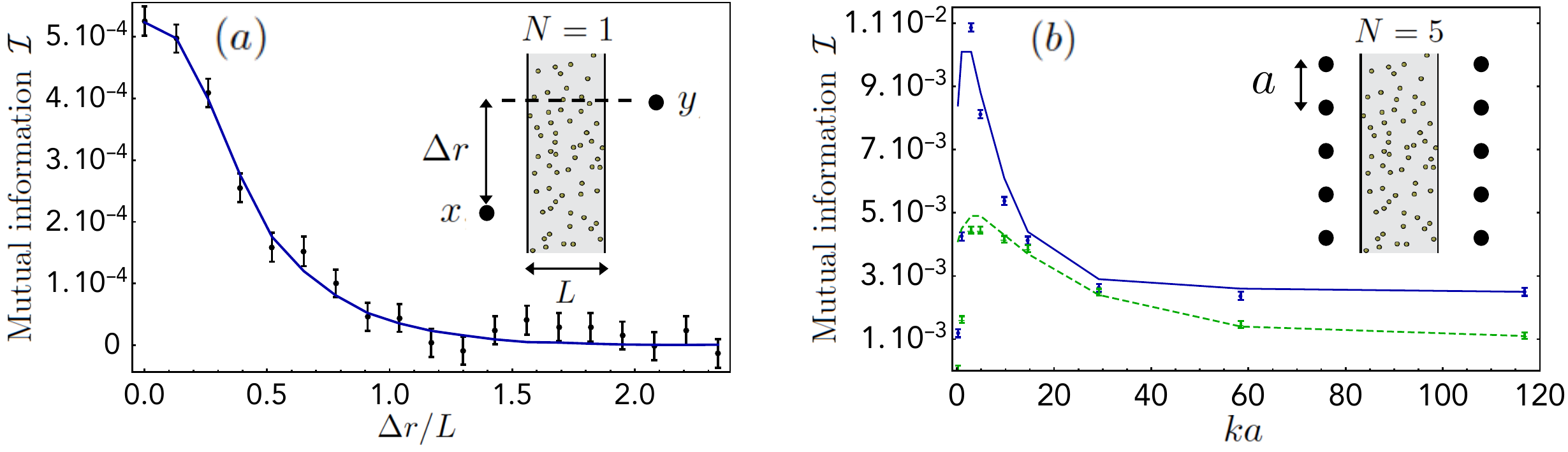}
\caption{MI as a function of distances between detectors.  Theoretical prediction (\ref{InfoTrace}) (lines) is compared to brute force estimation (dots) for various numbers of detectors $N$, thickness $L$, and mean free path $\ell$. (a) MI between the intensities measured in reflection ($x=I^R/\av{I^R}$) and transmission ($y=I^T/\av{I^T}$) vs transverse distance $\Delta r$. Parameters of the wave propagation simulation: $kL=30$, $k\ell=10$. (b)  MI between two sets of $N=5$ detectors vs detector spacing $a$, for two thicknesses, $kL=30$ (solid line) and $kL=80$ (dashed line), and fixed $k\ell=10$. Constant residual bias in numerical estimates (dots) have been removed, according to the procedure detailed in SI. Note that the agreement between simulations and theory for $ka\lesssim 1$ can only be qualitative since the hypothesis of weak surface correlations is not fully satisfied.} 
\label{fig2}
\end{figure*} 

In order to validate the theoretical prediction (\ref{InfoTrace}) or its approximation (\ref{InfoTrace2}), we have performed numerical simulations of wave propagation in two-dimensional (2D) disordered slabs with various thickness $L$ and scattering mean free path $\ell$. Subwavelength dipole scatterers were placed at random positions inside the slab, and the scalar wave equation was solved numerically using the coupled-dipole method \cite{SI}. For each set of parameters, $M=10^8$ disorder realizations were typically generated numerically, and the values of $\vec{x}$ and $\vec{y}$ were calculated at the sample input and output surfaces, for various numbers $N$ of detectors and interdistance $a$ between detectors. Then, from the sets of data $\{ \vec{x}_{\alpha},\vec{y}_{\alpha}\}_{\alpha=1 \dots M}$, an estimator of the MI was built, based on entropy estimates from nearest neighbor distances~\cite{SI, kraskov04}. Such an estimator is expected to be more accurate than binning estimators ---that consist in partitioning the support of $\vec{x}$ and $\vec{y}$ into bins --- for which the bias potentially grows exponentially with the dimension $N$ of $\vec{x}$ and $\vec{y}$~\cite{moddemeijer89}. 

Let us first analyze the simplest situation where a single pair of detection points is considered ($N=1$). The approximation (\ref{InfoTrace2}) takes in this case the simple form $\mc{I}\simeq C^{RT}(\Delta r)^2/2\textrm{ln}2$, where $\Delta r$ is the transverse distance between the detection points placed on both sides of the sample. As shown in Fig.~\ref{fig2} (a), this prediction agrees well with the direct estimate of Eq.~(\ref{def_info}), proving that the MI between $\vec{x}$ and $\vec{y}$  essentially boils down to the square of their correlation function for $N=1$. As a consequence, the MI is vanishingly small for $\Delta r\gtrsim L$.  Indeed, the correlation function $C^{RT}$ is transported along diffusive paths, that explore a transverse distance $\sim L$ in the multiple scattering regime $kL\gg k\ell \gg 1$, with $k=2\pi/\lambda$ \cite{starshynov17}. 

For a larger number of detectors ($N>1$), the behavior of the MI becomes more complex. Let us analyze its dependence on the interdistance $a$ between detectors. Results corresponding to samples with two different thicknesses $L$ are presented in Fig.~\ref{fig2} (b). Here also we obtain very good agreement between brute-force numerical estimates of the MI and the trace formula (\ref{InfoTrace}) completed by Eq.~(\ref{Cbar}), in which the values of correlators have been obtained from simulations.  This
confirms that MI in multiple scattering environments can be computed from the combination of pairwise correlators only. More precisely, we distinguish three regimes in Fig.~\ref{fig2}(b), that can be interpreted by means of the approximation (\ref{InfoTrace2}). For detector spacing $a$ larger than the extent $L$ of $C^{RT}(\Delta r)$, the MI is driven by detectors placed in front of each other only. Thus, it is independent on $a$ and $N$ times larger than the MI obtained with a single pair of detectors placed on opposite sides with $\Delta r=0$ [see Fig.~\ref{fig2}(a)]. When $a$ is progressively reduced, the MI starts to increase since more and more pairwise cross-sample correlations get activated. In the absence of correlations between the various components of $\vec{x}$ or $\vec{y}$, this increase would hold for arbitrary small spacing $a$.  However, we observe that the MI reaches a maximum for a certain critical distance below which it falls down, thereby revealing the effect of surface correlations. The latter contain both short-range and long-range contributions \cite{akkermans07}. Short-range contributions, responsible for the size $\lambda$ of speckle spots, explain the convergence of the MI towards its $N=1$ limit when $a\ll \lambda$. Indeed, the MI cannot be increased by adding detectors located in the same speckle spot. Nevertheless, the qualitative analysis based on Eq.~(\ref{InfoTrace2}) does not allow us to infer which contribution triggers the value of the critical distance, and to explain why the MI is globally reduced when the thickness of the medium increases.

\begin{figure*}[t]
\includegraphics[width=1\linewidth]{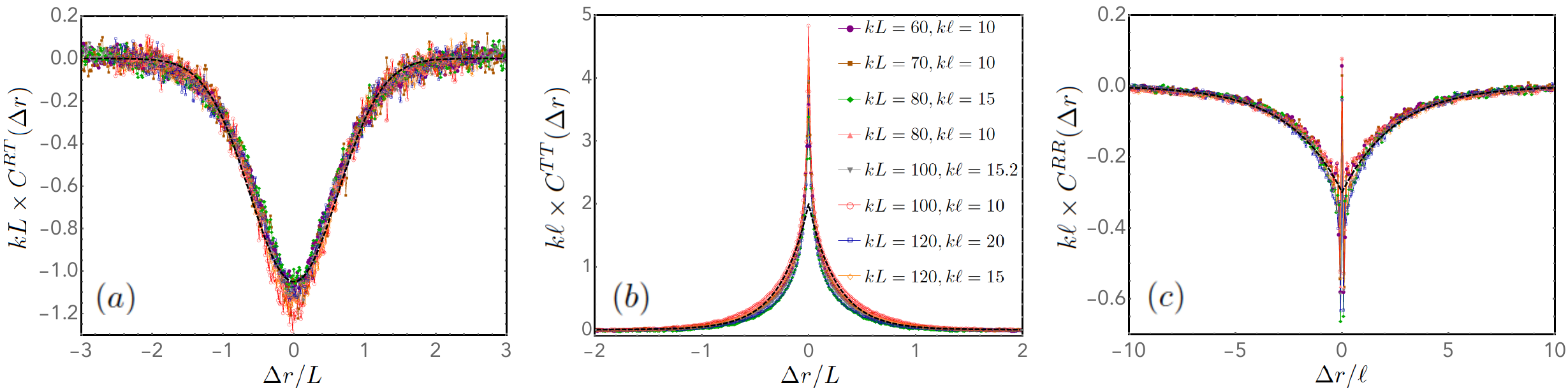}
\caption{Scaling of the three correlation functions that are building blocks of the MI: (a) $C^{RT}(\Delta r)$ (b) $C^{TT}(\Delta r)$ (c) $C^{RR}(\Delta r)$. Numerical results (dots) were obtained by solving the wave equation in 2D, for different values of $L$ and $\ell$ (see inset). Gaussian contributions ($C^{TT}_1$ and $C^{RR}_1$), which are short-range, have been removed for clarity. Dashed lines are simple fitting functions for $\Delta r \gtrsim \ell$ [gaussian in (a) and exponentials in (b) and (c)].}.
\label{fig3}
\end{figure*} 

To clarify these observations, we studied the dependence of the correlators $C^{RT}(\Delta r)$, $C^{TT}(\Delta r)$, and $C^{RR}(\Delta r)$ on $L$ and $\ell$. Simulation results for a plane wave illumination and various sets of parameters $\{kL, k\ell \}$ are shown in Fig.~\ref{fig3}. When properly normalized, data points collapse on singles curves, suggesting the following scalings in the regime $L\gg \ell \gg \lambda$: $C^{RT}(\Delta r)=-f_1(\Delta r/L)/(kL)^{d-1}$ for all $\Delta r$, and $C^{TT}(\Delta r)=(L/\ell)f_2(\Delta r/L)/(kL)^{d-1}$ and $C^{RR}(\Delta r)=-f_3(\Delta r/\ell)/(k\ell)^{d-1}$ for $\Delta r \gtrsim \ell$. Here $f_1$, $f_2$, and $f_3$ are three positive decaying function of range and amplitude close to unity, and $d$ is the space dimension. In the SI, we provide justifications for these scaling forms, based on random matrix theory and a microscopic diagrammatic approach. We highlight that, contrary to the well-established behavior of the long-range component of $C^{TT}$, $C^{RT}$ and $C^{RR}$ do not scale as $\sim 1/g$, where $g=k\ell(kL)^{d-2}$ is the Thouless number of a box of size $L$ \cite{stephen87,feng88, pnini89}. We also point out that the long-range component of $C^{RR}$ is negative, extending over a few mean free paths because waves explores such distance in the transverse direction before being reflected \cite{rogozkin95,froufe07}. Finally, it is instructive to note that the functions $f_1$, $f_2$, and $f_3$ are reasonably well fitted by a Gaussian and two exponentials (see Fig.~\ref{fig3}), which makes them useful in practical calculations.

\begin{figure}[b]
\includegraphics[width=0.9\linewidth]{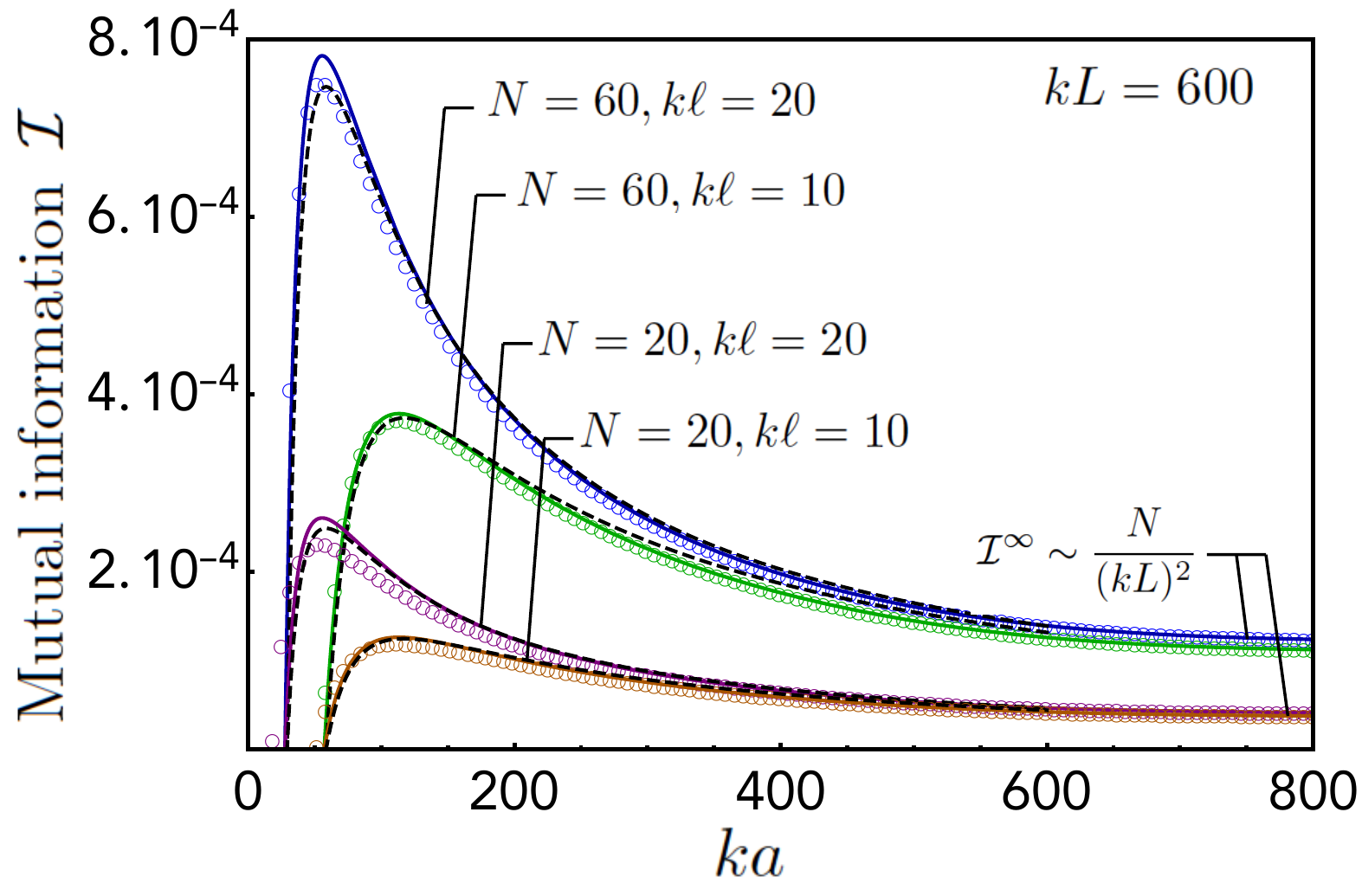}
\caption{Mutual Information computed from the trace formula (\ref{InfoTrace2}). Open circles are numerical calculations of the trace, using the scaling functions identified in Fig.~\ref{fig3} for $C^{RT}$,  $C^{TT}$, and $C^{RR}$. Solid lines stand for the analytic result presented in the SI, and dashed lines for its approximation (\ref{InfoContinuousLimit}) valid for $\ell \lesssim a \lesssim L$. }.
\label{fig4}
\end{figure} 

The  simple scaling forms of the three correlators allows us to push forward the analytic calculation of the trace formula (\ref{InfoTrace}), in particular in the interesting limit of a large number of detectors ($N\gg1$). As the three matrices $C^{RT}$, $C^{TT}$, and $C^{RR}$  are Toeplitz-type matrices, we may use an extension of the Szeg\"{o}'s theorem to evaluate the trace for arbitrary spacing $a$ \cite{gray06}. In order to simplify the discussion, we focus on the situation where detectors (or pixels) are equally spaced in all directions on the surface, in the regime $\ell \lesssim a \ll L$ (see the SI for a study in the general case). In this regime, the contribution of $C^{RR}$ is negligible, and the remaining sums over indices appearing in the development of the trace of the matrix product can be replaced by space integrals on the surface. The approximation (\ref{InfoTrace2}) becomes  \cite{SI}:
\be
\mc{I}\simeq\frac{N}{2\textrm{ln}2[(kL)(ka)]^{d-1}} \left(c_{RT}-\frac{c_{TT}}{(ka)^{d-1}} \frac{L}{\ell}\right),
\label{InfoContinuousLimit}
\ee
where $c_{RT}=\int \str{d}\vec{r}f_1(r)^2$ and $c_{TT}=\iint \str{d}\vec{r}\str{d}\vec{r'}f_1(r)f_2(r')f_1(\vert\vec{r}+\vec{r'}\vert)$ are two numerical constants on the order of unity. The result in Eq.~(\ref{InfoContinuousLimit}) supports previous qualitative observations: the MI scales linearly with the number of detectors, and it decreases when the sample thickness increases because the cross-sample correlation $C^{RT}$ itself is reduced. Interestingly, when we normalize Eq.~(\ref{InfoContinuousLimit}) by the MI measured for a single detector $\mc{I}_1=C^{RT}(0)^2/2\textrm{ln}2$, we obtain $\mc{I}/\mc{I}_1\varpropto N(c_{RT} u -c_{TT}u^2/g)$ where $u=(L/a)^{d-1}$. This shows that $\mc{I}$ exhibits a maximum triggered by the long-range component of $C^{TT}$, of the form $\mathcal{I}^{\str{max}} \sim N g \, \mc{I}_1$, for a critical interdistance $a^*$ much larger than the wavelength [$(L/a^*)^{d-1}=(c_{RT}/2c_{TT})g$ so that $a^* \sim \lambda (L/\ell)^{1/(d-1)}$]. Hence, the MI for an array of $N$ detectors with optimized interdistance is enhanced by a factor $Ng \gg 1$ compared to the MI for a single detector. These considerations are confirmed in Fig.~\ref{fig4} by the good agreement of the direct numerical evaluation of the trace $(\ref{InfoTrace2})$ with the full analytic prediction detailed in the SI and its approximation (\ref{InfoContinuousLimit}). In particular, denoting by $\mathcal{I}^\infty= N \mc{I}_1$ the MI obtained in the large spacing regime $a\gtrsim L$ where only frontside correlations contribute, we clearly observe the enhancement factor $\mathcal{I}^{\str{max}}/\mathcal{I}^\infty \sim g$.
 
In summary, we have presented a quantitative treatment of the MI between two speckle images produced on opposite sides of a multiple scattering medium. The dependence of the MI on length scales characterising the medium and on the detection geometry highlights the entangled and competitive contributions of long-range intensity correlations. In particular, we have shown that using an array of $N$ detectors with interdistance $a$ to record the speckle image, the MI can be increased by a factor of $Ng$ compared to the single detector case for a critical value of $a\gg \lambda$. Although our approach does not give the recipe to recover from $\vec{x}$ the information contained in $\vec{y}$, or \textit{vice-versa}, it provides quantitative estimates of the MI, and conditions for its optimization, that may be useful for the design of new setups dedicated to information recovery or transfer in complex media. 

This research was supported by LABEX WIFI (Laboratory of Excellence within the French
Program ÒInvestments for the FutureÓ) under references ANR-10-LABX-24 and ANR-10-IDEX-0001-02 PSL*. N.F. acknowledges financial support from the French ÒDirection G\'en\'erale de l'ArmementÓ (DGA).

\end{document}


\title{Supplementary Information \\  Mutual information between reflected and transmitted speckle images}

\author{
 N. Fayard, A. Goetschy, R. Pierrat, and R. Carminati,
}

\affiliation{
ESPCI Paris, PSL Research University, CNRS, Institut Langevin, 1 rue Jussieu, Paris, France
}

\maketitle

\section{I. Proof of the trace formula}

Our goal here is to prove Eqs.~(3),~(5), and~(6) of the main text (MT), and provide a generalization of Eq.~(5) that includes second-order corrections in $C_2^{TT}$. 

First we search for an expression of the jpdf $p(\vec{x},\vec{y})$ in terms of $p(\vec{x})$ and $p(\vec{y})$. This distribution is characterized by the set of correlators
\be
\av{x^{\{n\}}y^{\{m\}}}=
\av{
x_1^{n_1}...x_N^{n_N}y_1^{m_1}...y_N^{m_N}
\label{TotMomentum}
}.
\ee
As explained in the MT, we are interested in leading corrections to the independent result
$\av{x^{\{n\}}y^{\{m\}}}=\av{x^{\{n\}}} \av{y^{\{m\}}}$. We adopt a path-integral-type representation of each field
$E_i$ ($ E_i=\sum_{\mc{S}}E_i^{\mc{S}}$, where $\mc{S}$ is a scattering trajectory), such that each term
$x_i^{n_i}\varpropto \vert E_i^R \vert^{2n_i}$ contains $n_i$ replica of propagators $E_i^{R,\mc{S}}$, $E_i^{R,\mc{S}*}$. The
same representation is used for $y_j^{n_j} \varpropto \vert 
E_j^T \vert^{2n_j}$. We insert this decomposition into
Eq.~(\ref{TotMomentum}) and retain trajectories that provide non-zero contributions after averaging over scatterer
positions. Leading corrections to the independent result are due to correlations that involve propagator quadruplets,
such as $\{E_i^{R,\mc{S_1}}, E_i^{R,\mc{S_2}^*}, E_j^{T,\mc{S_3}}, E_j^{T,\mc{S_4}*}\}$. The number of these
quadruplets formed from replica is $n_i^2n_j^2$. In addition, their weight is $\av{\delta x_i \delta y_j}=\av{ x_i
y_j}-\av{ x_i }\av{ y_j}$. The correlator (\ref{TotMomentum}) becomes:
\begin{align}
\av{x^{\{n\}}y^{\{m\}}}&\simeq \av{x^{\{n\}}} \av{y^{\{m\}}} 
\nonumber
\\
&+\sum_{i,j}n_i^2n_j^2 \, \av{\delta x_i \delta y_j}\av{x^{\{n\}-1_i}}  \av{y^{\{m\}-1_j}},
\label{EqMoment}
\end{align}
where we use the notation $x^{\{n\}-1_i}=x_1^{n_1}\dots x^{n_i-1}\dots x_N^{n_N}$.  Then, we compute the characteristic function
\be
g(\vec{z},\vec{z'})=\av{e^{\vec{z}.\vec{x}+\vec{z'}.\vec{y}}}.
\ee 
By inserting the result (\ref{EqMoment}) into the series expansion of $g(\vec{z},\vec{z'})$ and using the property
\be
\sum_{n_i=0}^{\infty}\frac{n_iz_i^{n_i}}{(n_i-1)!}x^{\{n\}-1_i}= z_i \, \partial_{z_i}\left(
z_i \sum_{n_i=0}^{\infty} \frac{z_i^{n_i}}{n_i!}x^{\{n\}}
\right),
\ee
we get
\begin{align}
g(\vec{z},\vec{z'})&=g(\vec{z})g(\vec{z'})
\nonumber
\\
&+\sum_{i,j} \av{\delta x_i \delta y_j}\, z_i \, \partial_{z_i}\left[z_i g(\vec{z}) \right] z'_j \, \partial_{z'_j}\left[z'_j g(\vec{z'}) \right].
\end{align}
The distribution $p(\vec{x},\vec{y})$ follows by taking the inverse Laplace transform of $g(\vec{z},\vec{z'})$. It can be written in the following convenient form:
\be
p(\vec{x},\vec{y})=p(\vec{x})p(\vec{y})[1+u(\vec{x},\vec{y})],
\label{PofXY}
\ee
where the correction term $u(\vec{x},\vec{y})$ involves all possible pairwise correlations between the reflected and the transmitted speckles:
\begin{align}
u(\vec{x},\vec{y})&=\sum_{i,j}\av{\delta x_i \delta y_j} v_i(\vec{x}) v_j(\vec{y}),
\\
v_i(\vec{x})&=
\frac{\partial_{x_i}[x_i \partial_{x_i}p(\vec{x})]}
{p(\vec{x})}.
\label{DefVi}
\end{align}

We are now in position to compute explicitly the mutual information (MI). With the representation (\ref{PofXY}), it reads
\be
\mc{I}=\frac{1}{\textrm{ln}2}\iint \!\! \str{d}\vec{x}\str{d}\vec{y}p(\vec{x})p(\vec{y})[1+u(\vec{x},\vec{y})]\,\textrm{ln}\left[1+u(\vec{x},\vec{y})\right].
\label{InfoVsU}
\ee
We then expand the integrand as
\be
(1+u)\ln(1+u)=u+ \sum_{n=2}^{\infty}\frac{(-1)^{n}}{n(n-1)}u^n,
\ee
and note that the first term of this series does not contribute to (\ref{InfoVsU}). Indeed, using the notation $\av{\dots}_0=\iint \str{d}\vec{x}\str{d}\vec{y}p(\vec{x})p(\vec{y}) (\dots)$, we obtain
\begin{align}
\av{u(\vec{x},\vec{y})}_0 
&=\sum_{i,j}\av{\delta x_i \delta y_j} \av{v_i(\vec{x})}_0 \av{ v_j(\vec{y})}_0
\nonumber
\\
&=0,
\end{align}
since $\int \str{d}\vec{x} \, \partial_{x_i}[x_i \partial_{x_i}p(\vec{x})]=0$. 
This result holds whatever the distributions $p(\vec{x})$ and $p(\vec{y})$. Therefore, MI reduces to
\be
\mc{I}=\sum_{n=2}^{\infty}\frac{(-1)^{n}}{n(n-1)} \av{u(\vec{x},\vec{y})^n}_0.
\label{InfoVsU2}
\ee
Inasmuch as the images $\vec{x}$ and $\vec{y}$ are weakly correlated, we retain in the expansion (\ref{InfoVsU2}) the first term only. The term $ \av{u(\vec{x},\vec{y})^2}_0$ gives rise to the trace formula:
\be
\mc{I}\simeq\frac{1}{2\textrm{ln}2}\textrm{Tr}\left[
C^{RT}\bar{C}^{TT}C^{RT}\bar{C}^{RR}
\right],
\label{InfoTrace}
\ee
where we introduced three $N\times N$ matrices defined by their elements, $C^{RT}_{ij}=\av{\delta x_i\delta y_j}$
,
\begin{align}
\bar{C}^{TT}_{jj'}&=\av{ v_j (\vec{y})  v_{j'}(\vec{y})}_0
\label{CbarmatTT}
\\
&=\int \str{d}\vec{y} \, \frac{\partial_{y_j}[y_j \partial_{y_j}p(\vec{y})]\partial_{y_{j'}}[y_{j'} \partial_{y_{j'}}p(\vec{y})]}{p(\vec{y})},
\nonumber
\\
\bar{C}^{RR}_{i'i}&=\av{ v_{i'} (\vec{x})  v_{i}(\vec{x})}_0
\label{CbarmatRR}
\\
&=\int \str{d}\vec{x} \,  \frac{\partial_{x_{i'}}[x_{i'} \partial_{x_{i'}}p(\vec{x})]\partial_{x_{i}}[x_{i} \partial_{x_{i}}p(\vec{x})]}{p(\vec{x})}.
\nonumber
\end{align}

Next, we assume that the distance $a$ between consecutive pixels or detectors used to record the images is larger than
the wavelength $\lambda$. In this case, only weak but long-range parts of pairwise correlations inside each image
contribute to the jpdf $p(\vec{y})$ and $p(\vec{x})$. In order to find explicit forms for these distributions, we
proceed in the same way as for the moments (\ref{TotMomentum}), \textit{i.e.} we consider in the expansion of
$\av{y^{\{n\}}}$ (or $\av{x^{\{n\}}}$) all corrections due to the correlation of field quadruplets. We get an expansion
similar to Eq.~(\ref{EqMoment}):
\begin{align}
\av{y^{\{n\}}}= \prod_j\av{y^{n_j}} + \sum_{j<j'} & n_j^2n_{j'}^2 \, \av{\delta y_j \delta y_{j'}}\av{y^{n_j-1}}  \av{y^{n_{j'}-1}}
\nonumber
\\
& \times \prod_{k\neq j, k'\neq j' } \av{y_k^{n_k}}\av{y_{k'}^{n_{k'}}}.
\label{EqMoment2}
\end{align}
Note that the sum runs over indices $j<j'$ instead of $i,j$ in Eq.~(\ref{EqMoment}) to avoid redundant counting of pair
correlations in the same image. The corresponding distribution, obtained by computing the inverse Laplace transform of
the characteristic function, has the same structure as Eq.~(\ref{PofXY}):
\be
p(\vec{y})=\prod_k p(y_k) \left[ 1+ \sum_{j<j'} \av{\delta y_j \delta y_{j'}} v_j(y_j) v_{j'}(y_{j'}) \right],
\label{ApproxPofY}
\ee
where the definiton of $v_j(y_j)$ follows from Eq.~(\ref{DefVi}):
\be
\label{DefVj}
v_j(y_j)=\frac{\partial_{y_j}[y_j \partial_{y_i}p(y_j)]}{p(y_j)}.
\ee

The remaining unkown quantity in Eq.~(\ref{ApproxPofY}) is the distribution of intensity recorded on a single detector,
$p(y)$. This distribution has been calculated exactly in the nineties using random matrix theory as well as microscopic
diagrammatic approaches~\cite{kogan95, nieuwenhuizen95}. Here, we are interested in a tractable approximation of this distribution, that includes second-order corrections in $C_2^{TT}\simeq \av{(\delta y)^2}-1$.  For this purpose, we follow
 the approach of Ref.~\onlinecite{kogan93}, which is similar in spirit to the counting procedure used previously to evaluate the moments.   The moment $\av{y^n}$ are given by 
 \begin{align}
 \av{y^n}&\simeq \sum_{k=0}^{\str{E}(n/2)} \left(\frac{C_2^{TT}}{2}\right)^k  \,N_{n,k},
 \\
 &=(n!)^2\sum_{k=0}^{\str{E}(n/2)}
 \frac{\left(C_2^{TT}\right)^k}{4^kk!(n-2k)!}.
 \label{MomentY}
 \end{align} 
Here $C_2^{TT}$ is the weight of non-Gaussian corrections to the Rayleigh statistics, due to correlations of field quadruplets in the form of Hikami boxes. In addition, $N_{n,k}$ is the number of field combinations that contain $k$ Hikami boxes and $n-2k$ pairs of fields that form diffusons. In the following we keep terms in Eq.~(\ref{MomentY}) up to the second-order ($k\le 2$), so that the characteristic function $g(z)=\av{e^{zy}}$ reads
\be
g(z)=\frac{1}{1-z}+\frac{ z^2}{2(1-z)^3}C_2^{TT}+\frac{3 z^4}{4(1-z)^5}(C_2^{TT})^2,
\ee
and the distribution $p(y)$ becomes
\be
p(y)=e^{-y}\left[1+ h_1(y)C_2^{TT} +  h_2(y)(C_2^{TT})^2 \right],
\label{SolPofy}
\ee
with $h_1(y) = (y^2-4y+2)/4$ and $h_2(y) = (y^4-16y^3+72y^2-96y+24)/32$. Hence, the coefficient given by
Eq.~(\ref{DefVj}) and appearing in Eq.~(\ref{ApproxPofY}) is explicitly given by
\be
v_j(y_j)=\delta y_j+h_3 (y_j)C_2^{TT}+h_4(y_j)(C_2^{TT})^2,
\label{SolVofy}
\ee 
with $\delta y_j= y_j-1$, $h_3(y) =-y^2+3y-1$ and $h_4(y) =(7y^3-39y^2+50y-10)/4$. The jpdf $p(\vec{x})$ has a functional form identical to Eq.~(\ref{ApproxPofY}) with $C_2^{RR}$ replacing $C_2^{TT}$ in Eq.~(\ref{SolPofy}).

Finally, we compute the matrix elements (\ref{CbarmatTT}) and (\ref{CbarmatRR}), using Eqs.~(\ref{ApproxPofY}),
(\ref{SolPofy}) and (\ref{SolVofy}). This is a tedious but straightfoward calculation. For example, the first-order
expansion in $C_2^{TT}$ and $\av{\delta y_j \delta y_{\alpha}}$ of the coefficient given by Eq.~(\ref{DefVi}) that enters into the definitions  (\ref{CbarmatTT}) and (\ref{CbarmatRR}) reads
\begin{align}
&v_j (\vec{y})=\delta y_j +(1-2y_j) \sum_{\alpha \neq j } \av{\delta y_j \delta y_\alpha} \delta y_\alpha
+
h_3(y_j)C_2^{TT} 
\nonumber
\\
&+ \sum_{\alpha \neq j } \av{\delta y_j \delta y_\alpha} [h_5(y_j)\delta y_\alpha+(1-2y_j)h_3(y_\alpha)]
C_2^{TT},
\end{align}
with $h_5(y)=5y^2-12y+3$. The final result, obtained after integration over $\vec{y}$, looks quite simple:
\begin{align}
\bar{C}^{TT}_{jj'}&=\left[1-C_2^{TT}+5\left(C_2^{TT}\right)^2\right]\delta_{jj'}
\nonumber
\\
&-(1-2C_2^{TT})\av{\delta y_j \delta y_{j'}}(1-\delta_{jj'}).
\label{CbarFinal}
\end{align}
This completes the proof of Eq.~(5) of the MT, including second-order correction in $C_2^{TT}$. We note that second-order correction shows up in the diagonal matrix element only. We also stress that the result (\ref{CbarFinal}) is a first order expansion in $\av{\delta y_j \delta y_{j'}}$ ($j\neq j'$), as the distribution (\ref{ApproxPofY}) itself. Numerical simulations confirm that this approximation is sufficient for $a\gtrsim \lambda$.

\section{II. Wave equation simulations with the couple dipole methods}

The numerical simulations were performed using the coupled-dipole method~\cite{lax52}. This is an exact method providing
that the scatterers can be described by dipoles of subwavelength dimension. Here we considered cylindrical scatterers of
subwavelength cross-sections illuminated by an incident field polarized along their longitudinal axis, so that the wave
equation is effectively two dimensional. We first generate a configuration of the disorder by placing randomly $N_c$
cylinders in a box, the longitudinal size of which is given by the thickness $L$. The other dimension is
chosen to be large enough to mimic a slab geometry ($10$ times the thickness $L$). To avoid scatterers overlaps, a
minimum distance is forced between them. This distance is small enough not to introduce disorder correlations. Moreover,
to lower the number of scatterers inside the system and save computational time, the polarizability $\alpha$ of each
scatterer has been chosen such that it maximizes the scattering cross-section $\sigma_s=k^3|\alpha|^2/4$ while
verifying energy conservation [i.e. the scattering cross-section $\sigma_s$ should be equal to the extinction
cross-section $\sigma_e=k\, \operatorname{Im} (\alpha)$]. This leads to $\alpha=4i/k^2$. By adjusting the number
density of scatterers $\rho$, we can simulate systems with different scattering mean-free paths $\ell=1/(\rho\sigma_s)$.
The multiple interactions between the scatterers are described by a set of $N_c$ linear equations which reads
\begin{equation}
   E_j=E_0(\vec{r}_j)+k^2\alpha \sum_{n=1,n\ne j}^{N_c} G_0(\vec{r}_j-\vec{r}_n)E_n
\end{equation}
where $E_j$ is the exciting electric field on scatterer $j$ lying at position $\vec{r}_j$. $E_0$ is the
incident field (plane-wave at normal incidence) and $G_0(\vec{r}-\vec{r}_0)$ is the Green function in vacuum
which gives the electric field produced at position $\vec{r}$ by a source dipole lying at position $\vec{r}_0$. Its expression  reads
\begin{equation}
   G_0(\vec{r}-\vec{r}_0)=\frac{\textrm{i}}{4}H_0 (k\vert\vec{r}-\vec{r}_0\vert),
\end{equation}
where $H_0$ is the Hankel function of the first kind.  Once this system
is solved and the exciting fields are known, a similar equation is used to compute the field at any position at the sample surfaces:
\begin{equation}
   E(\vec{r})=E_0(\vec{r})+k^2\alpha\sum_{n=1}^{N_c} G_0(\vec{r}-\vec{r}_n)E_n.
\end{equation}
Finally, disorder averages are performed to compute the speckle correlation function $\langle
C^{RT}(\Delta\vec{r})\rangle$, $\langle C^{TT}(\Delta\vec{r})\rangle$, and $\langle C^{RR}(\Delta\vec{r})\rangle$. These correlation function are defined as
\be
C^{XY}(\Delta\vec{r})=\frac{\av{I^X(\vec{r})I^Y(\vec{r'})}}{\av{I^X(\vec{r})}  \av{I^Y(\vec{r'})}}-1,
\label{Def_correlation}
\ee
where $\bold{r'}=\bold{r} + \Delta \bold{r}$, the superscripts $X$ and $Y$ stand for $R$ or $T$, and the intensities are defined as $I^X(\vec{r})=\vert \delta E^X(\vec{r}) \vert^2$, with $\delta E^X(\vec{r})=E^X(\vec{r})-\av{E^X(\vec{r})}$.
As an example, for  disorder strength $k\ell=10$ and sample thickness
$kL=100$, we  used $N_c=1750$ dipoles and  $6.4$ millions of configurations.

\section{III. Mutual information estimation}

\begin{figure}[t]
\includegraphics[width=0.95\linewidth]{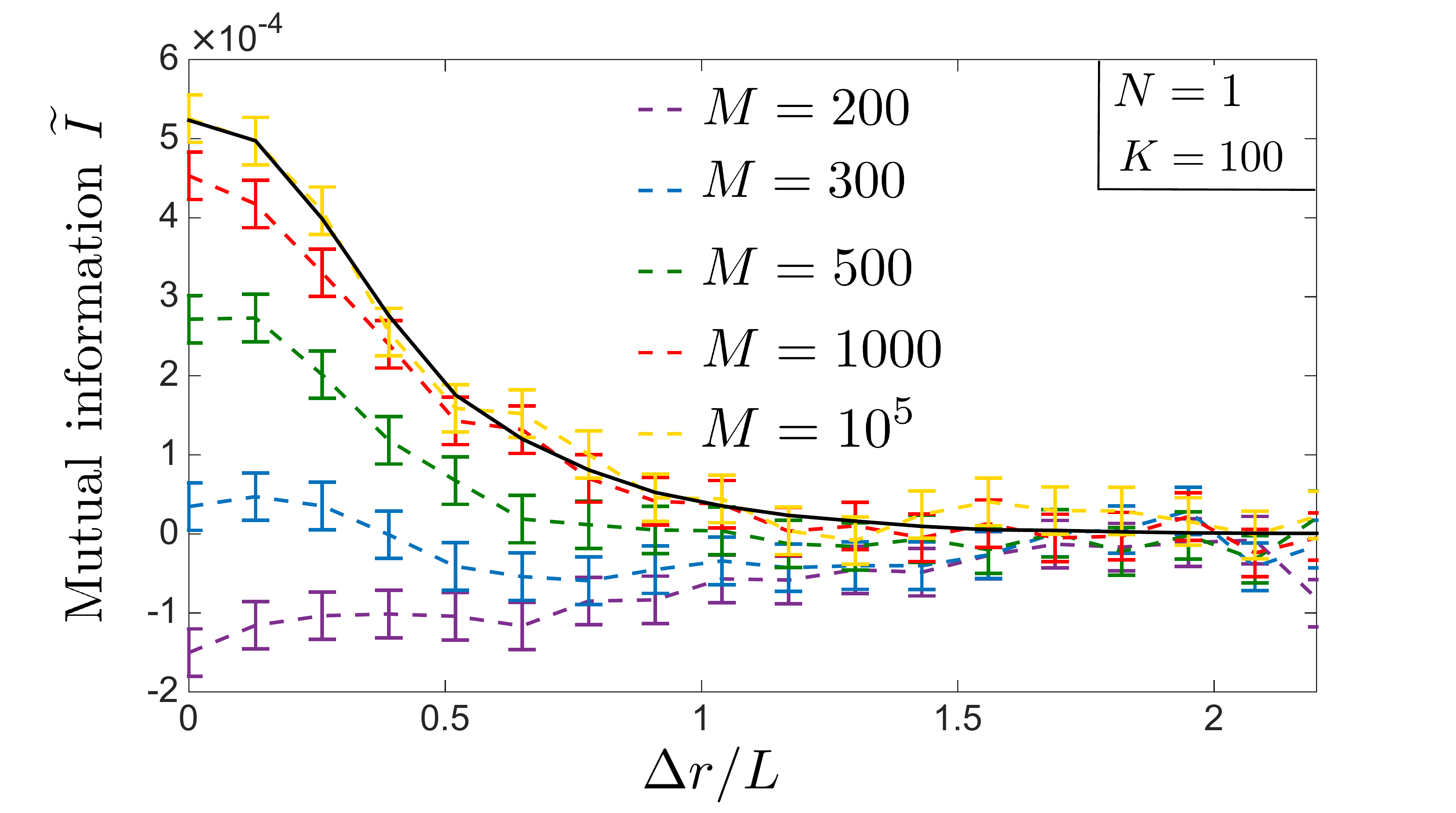}
\caption{MI between two detectors as function of their transverse distance $\Delta r$. Numerical estimates (dashed
lines) are computed from Eq.~(\ref{EstimatorFinal}) and compared to the theoretical result $\mathcal{I}=C^{RT}(\Delta
r)^2/2\ln{2}$ (solid line).}
\label{Fig1}
\end{figure}
 
Most intuitive estimators of  MI between $\vec{x}$ and $\vec{y}$ are binning estimators, that consist in partitioning the supports of $\vec{x}$ and $\vec{y}$ into bins, representing jpdf by histograms built from $M$ realizations, and approximating MI by a finite sum. Generally, these estimators suffer from bias due to finite $M$ and finite bin size~\cite{moddemeijer89}. In dimension $N=1$, it is possible to find a bin size that minimizes the bias, but no such strategy is available for the case $N>1$. In particular, bias remain non-zero in the limit $M\to \infty$ and it grows drastically with the dimension $N$. 

\begin{figure*}[t]
\includegraphics[width=0.95\linewidth]{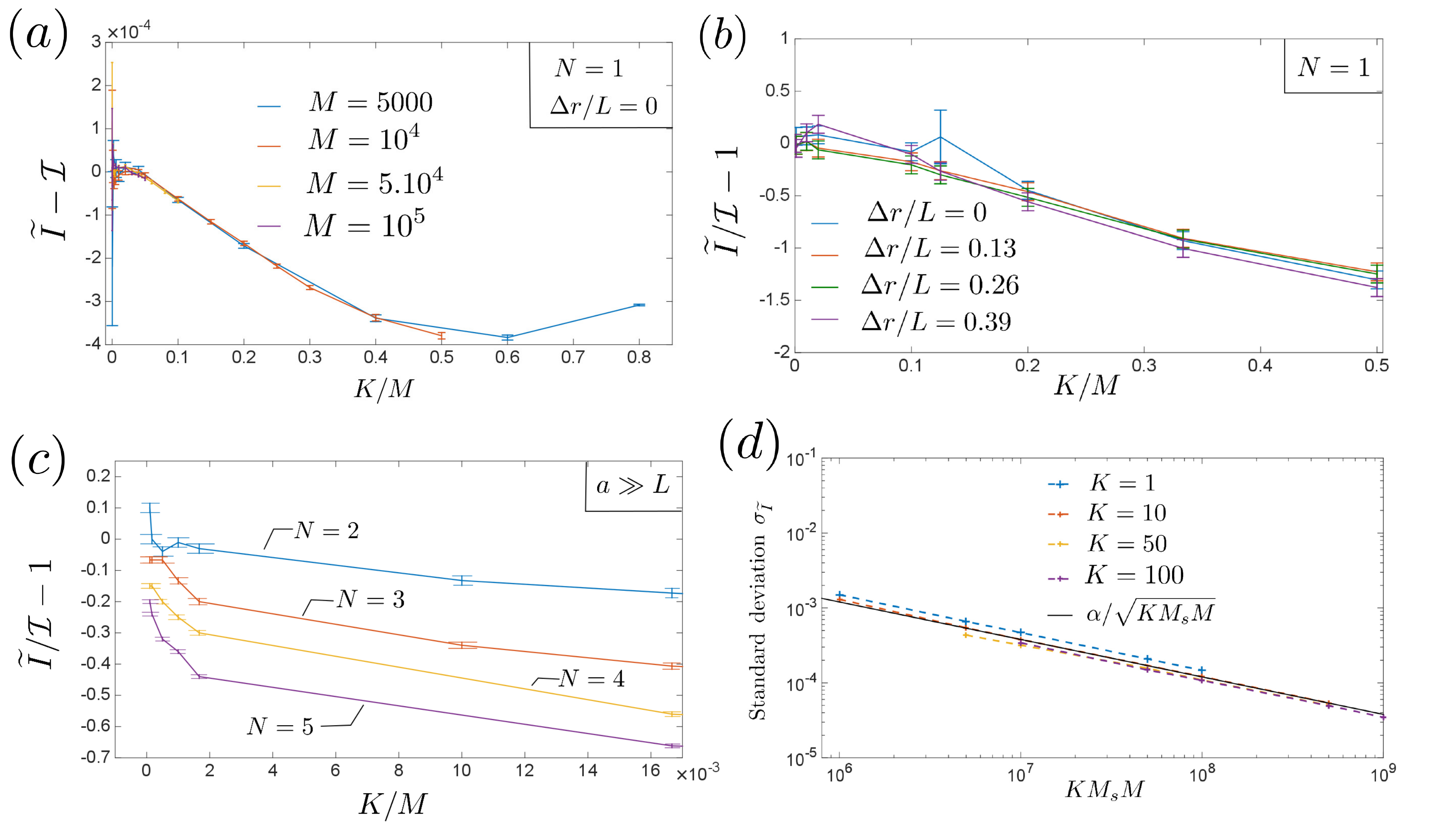}
\caption{Dependence of the bias $\tilde{\mathcal{I}}-\mathcal{I}$ on the number of realizations $M$ and neighbors
$K$ (a), on the MI $\mathcal{I}$ that decreases with $\Delta r$  (b), and on the numbers of detectors $N$ (c). Scaling
of the standard deviation of the estimator given in Eq.~(\ref{EstimatorFinal}) (d). In all panels, the estimator $\tilde{\mathcal{I}}$ has been averaged over $M_s$ sets of $M$ realizations to reduce the standard deviation. Parameters of the wave propagation
simulation are $kL=30$ and $k\ell=10$.}
\label{Fig2}
\end{figure*}
In order to limit the previous bias issue, we used an estimator of MI based on entropy estimates built from nearest
neighbor distances, measured in the space spanned by $\vec{x}$ and $\vec{y}$ ~\cite{kraskov04}. In the following, we
briefly summarize the approach of Ref.~\cite{kraskov04}, with the aim of clarifying the bias dependence on the system
parameters.  First we interpret the entropy $\mathcal{H}(\bold{x},\bold{y})$ as the average of
$\operatorname{log}p(\vec{z})$, with $\vec{z}=(\bold{x}, \bold{y})$. Its unbiased estimator, built from the data set
$\{\vec{z}_\alpha\}_{\alpha=1\dots M}$, is $\widetilde{\mathcal{H}}(\bold{x},\bold{y})=
-\overline{\operatorname{log}p(\bold{z}_\alpha)}$, where we use the notation
$\overline{(\dots)_{\alpha}}=M^{-1}\sum_{\alpha=1}^M(\dots)_{\alpha}$. Then, we construct an estimate of
$\operatorname{log}p(\bold{z}_\alpha)$ by considering the ball centered in $\bold{z}_\alpha$ that contains its $K$
nearest neighbors. Let us denote by $\epsilon_\alpha$ the diameter of this ball. If it is small enough, we approximately have  $p(\bold{z}_\alpha)\varpropto p_\alpha(\epsilon_\alpha)/\epsilon_\alpha^{2N}$, where $p_\alpha(\epsilon)$ is the probability to get a realization in the ball of diameter $\epsilon$ centered at $\bold{z}_\alpha$. Up to an irrelevant constant, we obtain
 \be
\widetilde{\mathcal{H}}(\bold{x},\bold{y}) \simeq -\overline{\operatorname{log}p_\alpha(\epsilon_\alpha)}+2N\,\overline{\operatorname{log}\epsilon_\alpha}.
 \ee
This representation is useful because we can replace the first term by its average $\int \str{d}\epsilon P_K(\epsilon)\operatorname{log}p(\epsilon)$, where $P_K(\epsilon)$ is the probability to find a ball of diameter $\epsilon$ containing $K$ realizations. By expressing $P_K(\epsilon)$ in terms of $p(\epsilon)$, we easily compute the previous integral to obtain 
 \be
\widetilde{\mathcal{H}}(\bold{x},\bold{y}) \simeq  -\psi(K)+\psi(M)+2N\,\overline{\operatorname{log}\epsilon_\alpha}
\label{Estimator1},
\ee
where $\psi$ is the digamma function. Formula (\ref{Estimator1}) is true, in principle, for any value of $K$. However, $\epsilon_\alpha$ increases with $K$, so that previous approximations may break down at large $K$, resulting in a large bias (see below for illustration).

The same procedure can be adopted to construct an estimator of  $\mathcal{H}(\bold{x})$. The only subtle point is that $K$ has been defined in the space of dimension $2N$ spanned by $\vec{z}$, so that the effective number of neighbors in the marginal space of dimension $N$ is different from $K$. In particular, this number depends on the choice of the norm. In the following, we use the maximum norm $\vert \vec{z} \vert=\textrm{max}(\vert\vec{x}\vert, \vert\vec{y}\vert)$, so that $\epsilon=\textrm{max}(\epsilon^{\vec{x}}, \epsilon^{\vec{y}})$. In that case, the number of neighbors in the marginal space is approximately  $K^{\vec{x}}\simeq n^{\vec{x}}+1$, where $n^{\vec{x}}$ is the number of elements contained in the ball (defined in the marginal space) of diameter $\epsilon$. Hence, we have 
\be
\widetilde{\mathcal{H}}(\bold{x}) \simeq  -\overline{\psi(n^{\vec{x}}_\alpha+1)}+\psi(M)+N\,\overline{\operatorname{log}\epsilon_\alpha}.
\label{Estimator2}
\ee
By combining Eq.~(\ref{Estimator1}) and Eq.~(\ref{Estimator2}), we finally obtain an estimator of $\mathcal{I}=\mathcal{H}(\bold{x})+\mathcal{H}(\bold{y})-\mathcal{H}(\bold{x},\bold{y})$:
\be
\widetilde{\mathcal{I}}=\psi(K)+\psi(M)-\frac{1}{M}\sum_{\alpha=1}^M \left[\psi(n^{\vec{x}}_\alpha+1) +\psi(n^{\vec{y}}_\alpha+1) \right].
\label{EstimatorFinal}
\ee
An important property of this estimator is that its bias remains moderate even at large $N$ and it tends to zero when $M$ tends to infinity. 

In practice, we transform the data  $\{\vec{z}_\alpha\}_{\alpha=1\dots M}$ to make them almost uniformly distributed, in view of minimizing the bias. This is possible because MI is invariant under homeomorfic tranformation of the variables $\vec{x}$ and  $\vec{y}$. Using the fact that the marginal distributions of the components of $\vec{x}$ and  $\vec{y}$ are close to the Rayleigh distribution, we apply the transformation $x_i \to x'_i=e^{-x_i}$ and $y_j \to y'_j=e^{-y_j}$ for $i,j=1\dots N$. Then, we choose a value of $K$ (according to a strategy discussed below) and, for each $\vec{z'}_\alpha$, we search for $\epsilon_\alpha$ which is twice the distance to the $K$th neighbor of $\vec{z'}_\alpha$ in the sense of the maximal norm. Finally, we evaluate $n^{\vec{x'}}_\alpha$ and  $n^{\vec{y'}}_\alpha$ in the two marginal spaces and compute (\ref{EstimatorFinal}). 

\begin{figure*}[t]
\includegraphics[width=0.95\linewidth]{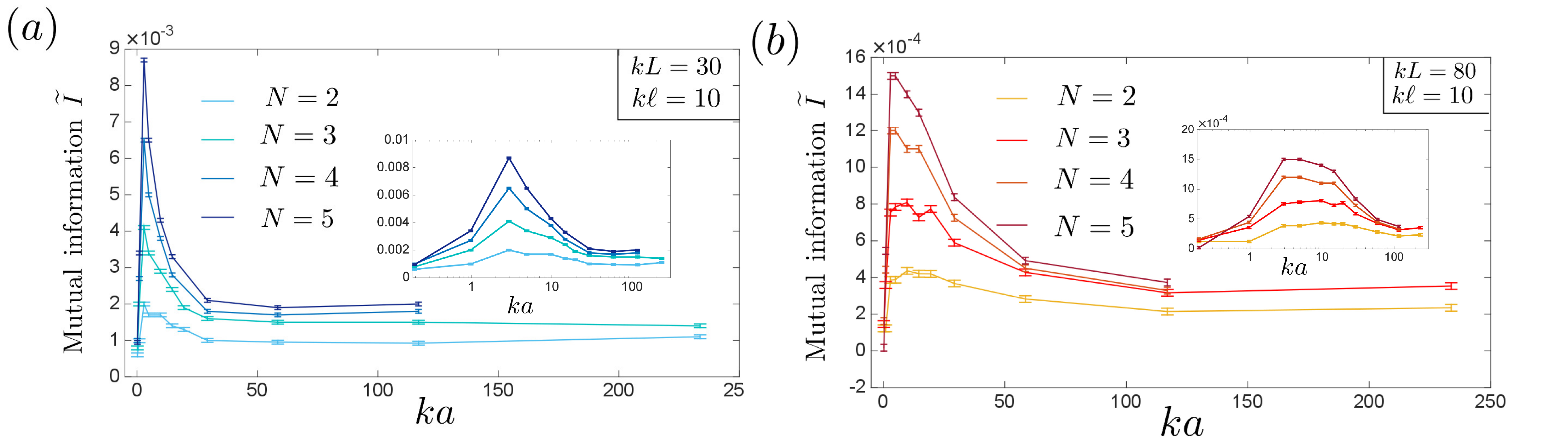}
\caption{MI between two sets of $N$ detectors vs detector spacing $a$, for two thicknesses $kL=30$ (a) and $kL=80$ (b),
and fixed $k\ell=10$.  Insets show the same data in log scale, in order to emphasize that the position and the amplitude
of the maximum are in agreement with theoretical predictions: $ka^*\varpropto L/\ell$ and
$\mathcal{I}^{\textrm{max}}/\mathcal{I}^\infty \simeq k\ell/4\simeq2.5$ ($c_{RT}\simeq c_{TT} \simeq 1.2$ in 2D, see
Sec.~V for details).} 
\label{Fig3}
\end{figure*}
In order to establish a reliable procedure for minimizing errors in estimates of MI, we studied the dependence of the the bias $\mathcal{B}=\av{\tilde{\mathcal{I}}}-\mathcal{I}$ and standard deviation $\sigma_{\tilde{\mathcal{I}}}$ of the estimator (\ref{EstimatorFinal}) on the different parameters of the problem:  number of realizations $M$,  number of nearest neighbors $K$,  value of the true MI $\mathcal{I}$, and number of detectors $N$. 
An illustration of the strong dependence of the bias on $M$ and $\mathcal{I}$ is given in Fig.~\ref{Fig1}, where we
represent  the same content as Fig.~2(a) of the MT --- \textit{i.e.} the MI between two detectors as a function of their
transverse distance --- for different values of $M$.  The comparison with the theoretical result
$\mathcal{I}=C^{RT}(\Delta r)^2/2\textrm{ln}2$ (solid lines) shows that the bias decreases at large $M$ and gets smaller
when we reduce $\mathcal{I}$ by increasing $\Delta r$. More precisely, the analysis of the bias for fixed values of $N$
and $\mathcal{I}$ reveals that the latter depends on the parameter $K/M$ only. This scaling dependence is shown in
Fig.~\ref{Fig2}(a) and agrees with similar analysis performed in Ref.~\onlinecite{kraskov04}. In addition, the bias
appears to be proportional to $\mathcal{I}$ [see Fig.~\ref{Fig2}(b)], so that it can be expressed in the form
\be
\mathcal{B}\simeq\mathcal{I}\operatorname{f}_N\left(\frac{K}{M}\right).
\label{ScalingB}
\ee
The function $\operatorname{f}_N$ is represented in Fig.~\ref{Fig2}(c) for various $N$.

We also studied the behavior of the standard deviation $\sigma_{\tilde{\mathcal{I}}}$ and found that it is independent of $\mathcal{I}$ and $N$ (data not shown), while it depends on $M$ and $K$ as $\sim (KM)^{-1/2}$. As expected from the construction of the estimator (\ref{EstimatorFinal}), $\sigma_{\tilde{\mathcal{I}}}$ decreases at large $K$ whereas the bias increases with $K$. We can further reduce $\sigma_{\tilde{\mathcal{I}}}$ without affecting significantly the computation time of the estimator by  averaging Eq.~(\ref{EstimatorFinal}) over $M_s$ sets of $M$ realizations. Hence, the standard deviation takes the form: 
\be
\sigma_{\tilde{\mathcal{I}}}\simeq\frac{\alpha}{\sqrt{KM_sM}},
\label{ScalingSigma}
\ee
where $\alpha\simeq 1.3$, according to Fig.~\ref{Fig2}(d).

In our simulations, we used the scaling forms (\ref{ScalingB}) and (\ref{ScalingSigma}) to infer the values of $K$ and
$M$ that are required to minimize errors in MI estimates. For a standard deviation satisfying
$\sigma_{\tilde{\mathcal{I}}}/\mathcal{I} \lesssim \epsilon$ where $\epsilon$ is some small number, we get $K\gtrsim
\alpha^2/M'\mathcal{I}^2\epsilon^2\simeq 10^{-8}/\mathcal{I}^2\epsilon^2$, since the product $M'=M_sM$ is kept fixed in
our study ($M'=10^8$). This gives us the value of $K$ required to estimate $\mathcal{I}$. Furthermore, a constraint on
the relative bias in the form $\mathcal{B}/\mathcal{I} \lesssim \epsilon$ translates into $f_N(k/M)\lesssim \epsilon$,
from which we find the constraint on $M$ using Fig.~\ref{Fig2}(c). Following this analysis, we chose the following
parameters in the MT: $K=10^2$, $M=10^5$ in Fig.~2(a) of the MT; $K=10$, $M=10^5$ for $kL=30$ and $K=10^2$, $M=10^6$ for
$kL=80$ in Fig.~2(b) of the MT.

Finally, we provide more details about Fig.~2(b) of the MT which focuses on the dependence of MI on the distance $a$ between detectors. We have represented in Fig.~(\ref{Fig3}) the MI computed from Eq.~(\ref{EstimatorFinal}) for various number $N$ of detectors. We observe that the estimate is not exactly linear with $N$, which is the signature of residual bias in the data. In order to get rid of them, we take advantage of Eq.~(\ref{ScalingB}) to express the true MI in the form
\begin{align}
\mathcal{I}&=\beta \tilde{\mathcal{I}},
\label{Rescale1}
\\
\beta&=\left[1+ \operatorname{f}_N(K/M)\right]^{-1}.
\end{align}
The proportionality coefficient $\beta$ depends on $N$ only. In particular, it is independent of the distance $a$, so that we can evaluate it in the large spacing regime $a\gg L$ where only frontside correlators contribute to MI:
\be
\beta=\frac{N\mathcal{I}_1}{\tilde{\mathcal{I}}^\infty}.
\label{Rescale2}
\ee
Figure 2(b) of the MT was obtained from Fig.~(\ref{Fig3}) by applying the rescaling given by Eqs.~(\ref{Rescale1}) and (\ref{Rescale2}),  with $\mathcal{I}_1$ deduced from Eqs. (\ref{InfoTrace})  and (\ref{CbarFinal}).

\section{IV. Scaling forms for the long-range components of $C^{RT}$, $C^{TT}$, and $C^{RR}$}

The correlations functions $C^{RT}$, $C^{TT}$, and $C^{RR}$ are defined according to Eq.~(\ref{Def_correlation}), where the intensities are square modulus of the fluctuating parts of the fields. In this way, we remove spurious interferences between mean fields and scattered fields, that vanish in the limit $L\gg\ell \gg \lambda$. The Gaussian contribution to Eq.~(\ref{Def_correlation}), denoted $C_1$, is obtained by pairing fields to form averages of complex conjugate pairs. On the other hand, non-Gaussian contributions necessarily involve scattering paths that connect four fields, since $\langle \delta E_X(\bold{r})\rangle=0$.  By noting $\langle \dots \rangle_c $ the non-Gaussian contributions and omitting the spatial dependence in the notation, we obtain
\begin{equation}
C^{XY}(\Delta \bold{r})=\frac{\vert\langle \delta E^X\delta E^{Y*}     \rangle\vert^2}{\av{I^X}  \av{I^Y}}+
\frac{\langle \delta E^X\delta E^{X*} \delta E^Y\delta E^{Y*}    \rangle_c}{\av{I^X}  \av{I^Y}}.
\label{Corr_decompo}
\end{equation}
The first term of Eq.~(\ref{Corr_decompo}) is the $C_1$ contribution. As far as $C^{TT}$ and $C^{RR}$ are concerned, this contribution is of the order of unity at short distances $\Delta\vec{r}\lesssim \lambda$. On the other hand, it is negligible in $C^{RT}$ for all distances as long as $L\gg \lambda$ \cite{starshynov17}. In the following, we disregard this well-known contribution and focus on the non-Gaussian term in Eq.~(\ref{Corr_decompo}). In the weak scattering regime $k\ell \gg1$, it is dominated by four field correlations made of one Hikami box, which are termed $C_2$. $C_2^{TT}$ and $C_2^{RR}$ contain both short-range and long-range components, while $C_2^{RT}$ does not contain any short-range feature. In what follows, we provide scaling forms for long-range components only, which we will label $C_{2, \textrm{long}}$. 

Two frameworks are available to compute $C^{XY}_{2, \textrm{long}}(\Delta \vec{r})$: the microscopic diagrammatic approach and random matrix theory (RMT) ~\cite{beenakker97, vanrossum99}. The diagrammatic formalism is in principle more powerful since it allows to compute all short-range and long-range features of the correlation functions. However, great care must be taken to properly account for leading contributions, in particular when reflected fields are involved~\cite{rogozkin95, rogozkin96}. To avoid such complications, we will mostly rely on the RMT approach which accounts, by construction, for all contributions ensuring flux conservation~\cite{mello90}. As we will show below, one major drawback is that it is based on an isotropy assumption that captures, in an open slab geometry, the integrated correlation  $\int C_{2, \textrm{long}}^{XY} (\Delta\vec{r})\textrm{d} \Delta\vec{r}$ only. This turns out to be sufficient to infer the scalings we are interested in.
\begin{figure*}[t]
\includegraphics[width=1\linewidth]{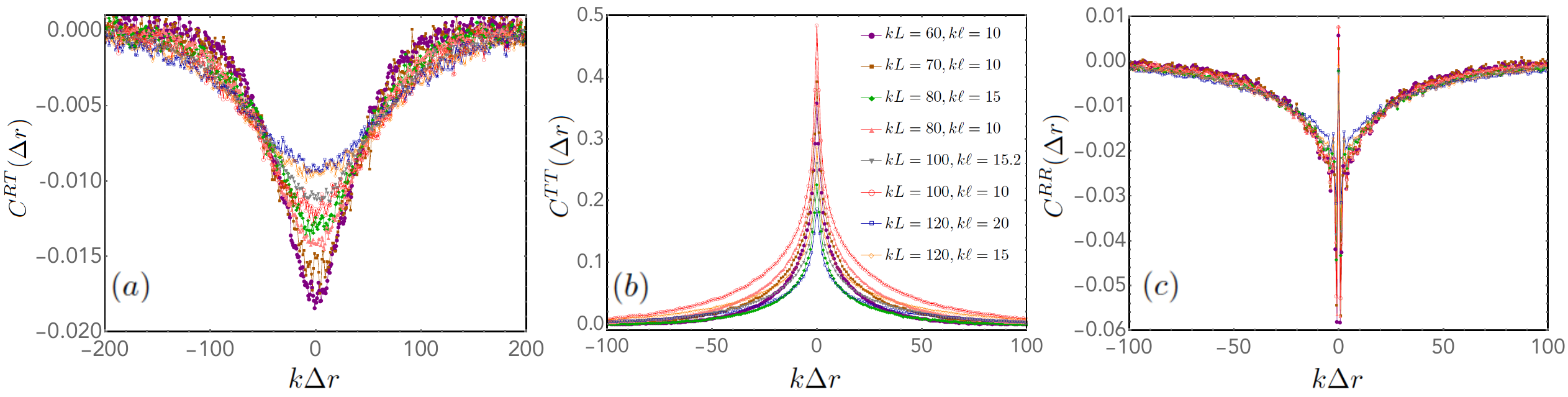}
\caption{Spatial dependence of the non-Gaussian part of the three correlation functions that build up the MI: (a) $C^{RT}(\Delta r)$ (b) $C^{TT}(\Delta r)$ (c) $C^{RR}(\Delta r)$. Numerical results were obtained by solving the wave equation in 2D, for  different values of $L$ and $\ell$ (see inset). The gaussian contribution to the correlation [first term of Eq.~(\ref{Corr_decompo})], which is short-range, has been removed for clarity.}.
\label{figCorr}
\end{figure*} 

With the help of the path-integral-type representation already used in Section I, we can show that far-field correlation
functions, $C^{XY}(\vec{k}_b,\vec{k}_{b'})$, are simply related to speckle correlation functions recorded at the sample
surface, $C^{XY}(\Delta\vec{r})$. In particular if we compute the correlators in the observation directions $\vec{k}_b=\vec{k}_{b'}$, $C_2^{XY}$ satisfy the following relations:
\begin{align}
C_2^{TT}(\vec{k}_b,\vec{k}_{b})&=2\int \frac{\textrm{d}\Delta \vec{r}}{\mathcal{A}}  C_{2, \textrm{long}}^{TT} (\Delta \bold{r}),
\label{CorrAngularSpace1}
\\
C_2^{RR}(\vec{k}_b,\vec{k}_{b})&=2\int \frac{\textrm{d}\Delta \vec{r}}{\mathcal{A}}  C_{2, \textrm{long}}^{RR} (\Delta \bold{r}),
\label{CorrAngularSpace2}
\\
C_2^{RT}(\vec{k}_b,\vec{k}_{b})&=\int \frac{\textrm{d}\Delta \vec{r}}{\mathcal{A}}  C_{2}^{RT} (\Delta \bold{r}),
\label{CorrAngularSpace3}
\end{align}
where $\mathcal{A}= W^{d-1}$ is the transverse area covered by the input illumination. Note the difference in the prefactors: there are two possibilities to form field pairing between the surface and the far-field for $C_2^{TT}$ and $C_2^{RR}$, while there is only one for $C_2^{RT}$ (the remaining pairing is negligible because it is made of mean fields crossing the full sample). 

Next, we evaluate the left hand-side of previous equations using RMT. Let us remind briefly the approach \cite{mello88, froufe07}. First we express the intensities $I^X(\vec{k}_b)$ as elements of transmission and reflection matrices of the slab, $t$ and $r$:  $I^T(\vec{k}_b)\varpropto \vert t_{ba}\vert^2$ and $I^R(\vec{k}_b)\varpropto \vert r_{ba}\vert^2$, where the subscript $a$ stands for the input plane wave $\vec{k}_a$. Second we use singular value decomposition of $t$ and $r$, taking into account constraints imposed by flux conservation and time-reversal symmetry:
\begin{align}
t&=U\sqrt{\tau}V^\dagger,
\\
r&=-V^*\sqrt{1-\tau}V^\dagger.
\end{align}
The RMT approach consists in assuming that $U$, $V$, and $\tau$ are three  independent  random matrices, with $U$ and $V$ uniformly distributed in the unitary group, and $\tau$ a diagonal matrix whose elements are the so-called transmission eigenvalues.
 The size of the matrices is equal to the number of propagating channels inside a waveguide of transverse section $\mathcal{A}$, $N_1=(kW/\pi)^{d-1}$. Using well-known statistical properties of random unitary matrices, we  first compute the leading contributions to the correlators $\av{I^X(\vec{k}_b)I^Y(\vec{k}_{b'})}$ in the limit $N_1\gg1$:
\begin{align}
\av{I^T(\vec{k}_b)I^T(\vec{k}_{b'})}& \varpropto\frac{1+\delta_{b,b'}}{N_1^4}\left[
\left(1-\frac{1}{N_1}\right)\av{\operatorname{Tr}^2(\tau)}\right.
\nonumber
\\
& \;\;\;\;\;\;\;\;\;\;\;\;\;\;\;\;\;\;\; +\left. \vphantom{\frac{1}{N}}\av{\operatorname{Tr}(\tau^2)}\right],
\label{Corr1}
\\
\av{I^R(\vec{k}_b)I^R(\vec{k}_{b'})}& \varpropto\frac{1+\delta_{b,b'}}{N_1^4}\left[
\left(1-\frac{2}{N_1}\right)\av{\operatorname{Tr}^2(1-\tau)}\right.
\nonumber
\\
& \;\;\;\;\;\;\;\;\;\;\;\;\;\;\;\;\;\;\; +\left. \vphantom{\frac{1}{N_1}}\av{\operatorname{Tr}(1-\tau)^2}\right],
\label{Corr2}
\\
\av{I^R(\vec{k}_b)I^T(\vec{k}_{b'})}& \varpropto\frac{1+\delta_{b,b'}}{N_1^4}\left[
\left(1-\frac{1}{N_1}\right)\av{\operatorname{Tr}(\tau)\operatorname{Tr}(1-\tau)}\right.
\nonumber
\\
& \;\;\;\;\;\;\;\;\;\;\;\;\;\;\;\;\;\;\; +\left. \vphantom{\frac{1}{N}}\av{\operatorname{Tr}(\tau(1-\tau)}\right].
\label{Corr3}
\end{align}
We stress that the RMT approach does not capture properly the dependence of these correlators in $\vec{k}_b-\vec{k}_{b'}$,  
because it assumes a quasi-one-dimensional geometry  ($W\ll L$), in which the density of states filters out wave-vector components $\vec{k}_b\neq\vec{k}_{b'}$. However, in the case $\vec{k}_b=\vec{k}_{b'}$, there is no filtering so that predictions from RMT are valid in the slab geometry as well ($W\gg L$).

From Eqs.~(\ref{Corr1}), (\ref{Corr2}), and (\ref{Corr3}), we can evaluate the correlation functions
$C^{XY}(\vec{k}_b,\vec{k}_{b'})$ explicitly, using $\av{\operatorname{Tr}(\tau^2)}=2N_1\bar{\tau}/3$, where  $\bar{\tau}\varpropto \ell/L$ is the mean transmission eigenvalue.  After having removed the $C_1$ contribution, we finally obtain
\begin{align}
C_2^{TT}(\vec{k}_b,\vec{k}_{b})&=\frac{4}{3}\frac{1}{N_1\bar{\tau}},
\\
C_2^{RR}(\vec{k}_b,\vec{k}_{b})&=\frac{2}{N_1}\left(-1+\frac{2\bar{\tau}}{3} \right),
\\
C_2^{RT}(\vec{k}_b,\vec{k}_{b})&=\frac{1}{N_1}\frac{-2/3+\bar{\tau}}{1-\bar{\tau}}.
\end{align}

We now have all the building blocks to evaluate the scaling forms of $C^{XY}_{2, \textrm{long}}(\Delta \vec{r})$. Combining previous formula  with  Eqs.~(\ref{CorrAngularSpace1}), (\ref{CorrAngularSpace2}), and (\ref{CorrAngularSpace3}), in the regime $\bar{\tau}\ll 1$, we find
\begin{align}
&\left(\frac{k}{\pi}\right)^{d-1}\!\!\int \textrm{d}\Delta \vec{r}  \;C_{2, \textrm{long}}^{TT} (\Delta \bold{r})=\frac{2}{3\bar{\tau}},
\label{CorrSpace1}
\\
&\left(\frac{k}{\pi}\right)^{d-1}\!\!\int \textrm{d}\Delta \vec{r}  \;C_{2, \textrm{long}}^{RR} (\Delta \bold{r})=-2,
\label{CorrSpace2}
\\
&\left(\frac{k}{\pi}\right)^{d-1}\!\!\int \textrm{d}\Delta \vec{r}  \;C_{2}^{RT} (\Delta \bold{r})=-\frac{2}{3}.
\label{CorrSpace3}
\end{align}
Typical transmitted waves, which cross the sample through a diffusion process, explore a transverse distance $\sim L$. On the other hand, typical reflected waves explore a transverse distance $\sim \ell$. For these reasons, we look for scaling functions in the form $C^{RT}(\Delta r)=\alpha_{RT}f_1(\Delta r/L)$, $C_{2, \textrm{long}}^{TT}(\Delta r)=\alpha_{TT}f_2(\Delta r/L)$, and $C_{2, \textrm{long}}^{RR}(\Delta r)=\alpha_{RR}f_3(\Delta r/\ell)$, where $f_1$, $f_2$, and $f_3$ are three positive decaying function of range and amplitude close to unity. Inserting these trials functions into Eqs.~(\ref{CorrSpace1}), (\ref{CorrSpace2}), and (\ref{CorrSpace3}), we find
\begin{align}
\alpha_{RT}&=-\frac{1}{(kL)^{d-1}},
\\
\alpha_{TT}&=\frac{1}{(kL)^{d-1}}\frac{L}{\ell},
\\
\alpha_{RR}&=-\frac{1}{(k\ell)^{d-1}},
\end{align}
where numerical prefactors have been absorbed in the definition of $f_1$, $f_2$, and $f_3$. Hence, we recover the scaling $C_{2, \textrm{long}}^{TT}\sim1/g$, with $g= k\ell (k L)^{d-2}$, which has been popularized in the eighties, as well as the scaling of $C^{RT}$ established recently with the help of the microscopic diagrammatic approach \cite{stephen87, feng88, pnini89, starshynov17}.

We confirmed in Fig.~3 of the MT the validity of previous scaling functions by showing that correlation functions,
computed numerically from wave equation simulation, collapse on single curves when properly normalized. For
completeness, we report in Fig.~\ref{figCorr} the same data without normalization.

\section{V. Analytic calculation of the trace formula}
We provide here an analytic evaluation of the trace formula (\ref{InfoTrace}),  in the form of the approximation (6) of the MT.  We reproduce it for convenience:
\be
\mc{I} \simeq
\frac{1}{2\textrm{ln}2}\textrm{Tr}\left[(\mathds{1}-{C}^{TT}-{C}^{RR})
(C^{RT})^2
\right].
\label{InfoTrace2}
\ee
The elements of the three $N \times N $ matrices under the trace are assumed to be of the following form:

\begin{align}
C^{RT}_{ij}&=-\frac{1}{(kL)^{d-1}}\,f_1\left(\vert \vec{r}_i -\vec{r}_{j} \vert/L \right),
\label{CRTMicro}
\\
C^{TT}_{jj'}&=\frac{1}{(kL)^{d-1}} \frac{L}{\ell}\,f_2\left(\vert \vec{r}_j -\vec{r}_{j'} \vert/L \right),
\label{CTTMicro}
\\
C^{RR}_{ii'}&=-\frac{1}{(k\ell)^{d-1}} \,f_3\left(\vert \vec{r}_i -\vec{r}_{i'} \vert/\ell \right),
\label{CRRMicro}
\end{align}
where $d$ is the space dimension and the functions $f_1$, $f_2$, and $f_3$ are three positive decaying functions of
amplitude and range close to unity (see Sec.~IV for justification of these scalings).
In the following, we focus for simplicity on the situation where detectors are separated by a distance $a$ in all directions of the surface.

 Let us start our analysis with the regime where the density of detectors is sufficiently large to approximate sums overs indices appearing in the development of Eq.~(\ref{InfoTrace2}) by integration over surface positions. According to the range of $f_1$,  $f_2$, and $f_3$, this regime requires  $a\lesssim L$ for $C^{RT}$ and $C^{TT}$, and $a\lesssim \ell$ for $C^{RR}$. In fact, this last condition may be relaxed since the contribution of $C^{RR}$ will turn out to be negligible for all values of $\lambda \lesssim a \lesssim L$ (see the discussion below). In this continous limit, we make the approximations:
\begin{align}
&\sum_{i,j}(C_{ij}^{RT})^2\simeq N \int \frac{\str{d}\mathbf{\Delta r}}{a^{d-1}}C^{RT}(\Delta r)^2,
\\
&\sum_{i,\alpha, \beta}C_{i\alpha}^{RT}C^{\textrm{XX}}_{\alpha \beta} C_{\beta i}^{RT}\simeq N \iint \frac{\str{d}\mathbf{\Delta r} \,\str{d}\mathbf{\Delta r'}}{a^{2(d-1)}}C^{RT}(\Delta r)
\nonumber
\\
&\;\;\;\;\;\;\;\;\;\;\;\;\;\;\;\;\;\;\;\;\;\;\;\;\;\;\;\;\;
\times C^{\textrm{XX}}(\Delta r')C^{RT}(\vert \mathbf{\Delta r} +  \mathbf{\Delta r'} \vert),
\end{align}
where the superscript $X$ stands for $R$ or $T$. Let us note $\mc{A}(W)$ the area of each image, parametrized by a typical transverse dimension $W$: $\mc{A}(W) \sim W^{d-1}$. With the scaling expressions (\ref{CRTMicro}),  (\ref{CTTMicro}), and (\ref{CRRMicro}), the mutual information (\ref{InfoTrace2}) becomes
\be
\mc{I}\simeq\frac{N}{2\textrm{ln}2[(kL)(ka)]^{d-1}} \left[c_{RT}-\frac{c_{TT}(L/\ell)-c_{RR}}{(ka)^{d-1}}\right],
\label{InfoContinuousLimit}
\ee
where we introduced the coefficients
\begin{align}
c_{RT}&=\int_{\mc{A}(W/L)} \str{d}\vec{r}f_1(r)^2,
\\
c_{TT}&=\iint_{\mc{A}(W/L)} \str{d}\vec{r}\str{d}\vec{r'}f_1(r)f_2(r')f_1(\vert\vec{r}+\vec{r'}\vert),
\\
c_{RR}&=\int_{\mc{A}(W/L)} \!\!\!\!\!\!\!\!   \str{d}\vec{r} \int_{\mc{A}(W/\ell)} \!\!\!\!\!\!\!\!\!\! \str{d}\vec{r'}f_1(r)f_3(r')f_1\left(\left\vert\vec{r}+(\ell/L)\vec{r'}\right\vert\right).
\end{align}
In the limit $W\gg L\gg \ell$, these coefficients  become constants of the order of unity and the result (\ref{InfoContinuousLimit}) reduces to Eq.~(7) of the MT.  For example, if we use the fitting functions $f_1(r)=\alpha_1e^{-\beta_1r^2}$ and $f_2(r)=\alpha_2e^{-\beta_2 r}$ as shown in Fig.~3 of MT, we get, in dimension $d=2$, $c_{RT}=\alpha_{1}^2\sqrt{\pi/2\beta_1}\simeq1.2$ and $c_{TT}=\pi\alpha_{1}^2\alpha_2e^{\beta_2^2/2\beta_1}[1-\erf(\beta_2/\sqrt{2\beta_1})]/\beta_1\simeq1.2$ ($\alpha_1\simeq1.05$, $\beta_1\simeq1.3$, $\alpha_2\simeq2$, and $\beta_2\simeq3.8$ were used to fit the data).

We now turn to the analytic evaluation of the trace (\ref{InfoTrace2}) without assuming any condition on $a$. This is
possible in the limit of large number of detectors ($N\gg 1$), where we can use powerful theorems for  spectral
properties of Toeplitz matrices \cite{gray06}. Furthermore, we restrict the analysis to the dimension $d=2$, where
numerical simulations have been performed  (see Sec.~II). In this case,  the matrices $C^{RT}$, $C^{TT}$, and $C^{RR}$ are standard Toeplitz matrices ($C_{ij}$ depends on $i-j$ only), while for $d=3$ we would have to deal with block Toeplitz matrices. Extension of the Szeg\"{o}'s theorem for products of Toeplitz matrices allows us to express the trace (\ref{InfoTrace2}) as
\be
\mc{I}=\frac{N}{2\textrm{ln}2}\int_0^{2\pi}\frac{ \str{d}\mu}{2\pi}\left[1- \tilde{C}^{TT}(\mu)-\tilde{C}^{RR}(\mu)\right]\tilde{C}^{RT}(\mu)^2,
\label{InfoToeplitz}
\ee
where 
\be
\tilde{C}^{XY}(\mu)=\sum_{n=-\infty}^{\infty}C^{XY}(a\vert n\vert)e^{\str{i} n \mu}.
\ee
Using Eqs.~(\ref{CRTMicro}),  (\ref{CTTMicro}), and (\ref{CRRMicro}) for $d=2$, we obtain 
\be
\mc{I}=\frac{N}{2\textrm{ln}2(kL)^{2}} \left(S^{RT}-\frac{S^{TT}-S^{RR}}{k\ell}\right),
\label{InfoToeplitz2}
\ee
where $S^{RT}$, $S^{TT}$, and $S^{RR}$ are three series defined as
\begin{align}
S^{RT}&= \sum_{n}f_1\left( \frac{\vert n \vert a}{L}\right)^2,
\\
S^{TT}&= \sum_{n,p}f_1\left( \frac{\vert n \vert a}{L}\right)f_1\left( \frac{\vert p \vert a}{L}\right)f_2\left( \frac{\vert n+p \vert a}{L}\right),
\label{SolSTT1}
\\
S^{RR}&= \sum_{n,p}f_1\left( \frac{\vert n \vert a}{L}\right)f_1\left( \frac{\vert p \vert a}{L}\right)f_3\left( \frac{\vert n+p \vert a}{\ell}\right).
\label{SolSRR1}
\end{align}
As discussed in the MT,  $f_1$ is well approximated by a Gaussian function, $f_1(r)=\alpha_1e^{-\beta_1 r^2}$, so that  $S^{RT}$ reduces to a Jacobi theta function:
\be
S^{RT}=\alpha_1^2 \, \theta\left[2\beta_1\left(\frac{a}{L}\right)^2\right].
\label{SolSRT}
\ee
 Here we use the notation $\theta(x) \equiv \theta_3(0, e^{-2x})$, where  $\theta_3(u, q)=1+2\sum_{n=1}^\infty q^{n^2}\cos(2nu)$. If we also use Gaussian models for $f_2$ and $f_3$  [$f_2(r)=\alpha_2 e^{-\beta_2 r^2}$ and $f_3(r)=\alpha_3 e^{-\beta_3 r^2}$], we find
 \begin{multline}
S^{TT}=\alpha_1^2\alpha_2 \, \theta\left[(\beta_1+\beta_2)\left(\frac{a}{L}\right)^2\right]
\\
\times \theta\left[\beta_1\frac{\beta_1+2\beta_2}{\beta_1+\beta_2}\left(\frac{a}{L}\right)^2\right].
 \end{multline}
and 
 \begin{multline}
S^{RR}=\alpha_1^2\alpha_3 \, \theta\left[\left(\beta_3+\beta_1\frac{\ell^2}{L^2}\right)\left(\frac{a}{\ell}\right)^2\right]
\\
\times \theta\left[\beta_1\frac{2\beta_3+\beta_1 \ell^2/L^2}{\beta_3+\beta_1\ell^2/L^2}\left(\frac{a}{L}\right)^2\right].
 \end{multline}
Simple approximations of these functions can be obtained using $\theta(x) \simeq\sqrt{\pi/x}$ for $x\ll 1$, and $\theta(x) \simeq1$ for $x\gg 1$. In this way, we can recover the result (\ref{InfoContinuousLimit}). We also find that $S^{TT} \sim (L/a)^2$ for $a\ll L $, $S^{RR} \sim L/a$ for $\ell \ll a \ll L $, and  $S^{RR} \sim (\ell/a)(L/a)$ for $a \ll \ell $. Hence, $S^{RR}\ll S^{TT}$ for $a\ll L $ in the multiple scattering regime $L \gg \ell$. This is the reason why we neglected the contribution of reflection correlations to mutual information in Eq.~(7) of the MT. In the opposite limit $a\gg L$, we find $S^{RT}\sim1$, $S^{TT}\sim1$, $S^{RR}\sim1$, so that surface correlations become negligible and $\mc{I}^\infty \sim N/(kL)^2$. 

Although Gaussian fits for $f_2(r)$ and $f_3(r)$ in the regime $r\gtrsim \ell$ are satisfactory, we used in the MT
exponential fits [$f_2(r)=\alpha_2 e^{-\beta_2 r}$ and $f_3(r)=\alpha_3 e^{-\beta_3 r}$] that give better agreement with
correlation functions computed from wave propagation simulations. The solid line in Fig.~4 of the MT represents the result (\ref{InfoToeplitz2}) with $S^{RT}$ given by Eq.~(\ref{SolSRT}), and $S^{TT}$ and $S^{RR}$ by Eqs.~(\ref{SolSTT1}) and (\ref{SolSRR1}) with exponential forms for $f_2$ and $f_3$.